# Bond type restricted radial distribution functions for accurate machine learning prediction of atomization energies

Mykhaylo Krykunov, Tom K. Woo

Department of Chemistry and Biomolecular Science
University of Ottawa, Ottawa, Canada

**Abstract**: Understanding the performance of machine learning algorithms is essential for designing more accurate and efficient statistical models. It is not always possible to unravel the reasoning of neural networks. Here we propose a method for calculating machine learning kernels in closed and analytic form by combining atomic property weighted radial distribution function (AP-RDF) descriptor with a Gaussian kernel. This allowed us to analyse and improve the performance of the Bag-of-Bonds descriptor, when the bond type restriction is included in AP-RDF. The improvement is achieved for the prediction of molecular atomization energies and is due to the incorporation of a tensor product into the kernel which captures the multidimensional representation of the AP-RDF. On the other hand, the numerical version of the AP-RDF is a constant size descriptor, and it is more computationally efficient than Bag-of-Bonds. We have also discussed a connection between molecular quantum similarity and machine learning kernels with first principles kind of descriptors.

## 1. Introduction

Machine learning is being applied to a broad range of areas including quantum chemistry where there is interest in replacing or augmenting traditional compute intensive quantum mechanical calculations with predictions based on molecular descriptors that can be rapidly calculated. A variety of descriptors have been developed for this purpose that have been used to predict many different properties, including the ground state energies of molecules and materials[1–3]. The accuracy of the results is quite sensitive to the descriptors used for predicting such energies for data sets of small organic molecules. Nonetheless, there has been a steady decrease in the mean average error (MAE) between the calculated DFT atomization energies and those predicted with machine learning.

One early descriptor used for predicting the atomization energies of small molecules is the Coulomb matrix, $\mathbf{C}$, which is defined as follows[1]

$$C_{ij} = \frac{Z_i Z_j}{|\mathbf{R}_i - \mathbf{R}_j|}, \tag{1}$$

$$C_{ii} = 0.5 Z_i^{2.4}. \tag{2}$$

It is notable that this descriptor contains the same information used in solving for the molecular electronic Schrödinger equation, namely, the nuclear positions, $\mathbf{R}_i$ and nuclear charges, $Z_i$. The drawback of this descriptor is that the **C** matrix is not invariant with respect to the permutation of atomic indices (the order of atoms). One way around this problem is to sort the matrix elements, with the largest elements always being placed at the top left position. This sorted Coulomb matrix was able to obtain a MAE of 4.3 kcal/mol in the atomization energy on a test set of 7165 small organic molecules composed of the elements C, O, N, S, and H and containing molecules up to 23 atoms in size[1]. Another approach explored to overcome the atom ordering dependence was to use the eigenvalues of the **C** matrix, which are invariant to the atomic indices permutations. However, this yielded a larger MAE of 10.0 kcal/mol[1] compared to the 4.3 kcal/mol[2] MAE obtained with the sorted Coulomb matrix. This is perhaps not surprising as the eigenvalues contain less information than the original Coulomb matrix. A third option, known as Bag-of-Bonds[3], places the Coulomb matrix elements into "bags" according to the chemical bond type. The Bag-of-Bonds method yielded MAE of the same test set ranging from 4.5 to 1.5 kcal/mol[3] depending on a machine learning kernel used. Descriptors that include additional information about bond angles and dihedral angles are even more accurate[4,5]. Similar approaches have been applied to periodic structures[6], but the level of accuracy seen with small molecules has not yet been achieved. This has been attributed to a huge compositional and structural diversity of the given dataset. Note that an increase of diversity in molecular systems also leads to a decrease of accuracy[7].

Periodic compounds such as metal organic frameworks (MOFs) are of particular interest to us. Such materials present another challenge in that it is not uncommon for these compounds to have unit cells with over a 1000 atoms. Thus, the descriptors in these cases should not only satisfy the requirement of accuracy, but they should also be manageable for statistical modeling. To clarify our point let us consider the Coulomb matrix of the size *N* by *N* of a chemical structure with *N* atoms. With compounds containing over a thousand atoms, the size of the Coulomb matrix would exceed one million. The Bag-of-Bonds descriptor has even more extended length, because the number of "bags" equals the number of chemical bond types in the whole database but not just in a given structure. Therefore, the total descriptor size would be again over a million.

Such a huge number of matrix elements would require a corresponding number of parameters for a statistical model. For example, methods that involve a linear transformation, i.e. $y = Ax + b$, require at least $M \times N + 1$ parameters for optimization, where *N* is the length of the descriptor vector $x$, and *M* equals one for linear ridge regression[8] or corresponds to the size of a subsequent layer in feedforward neural networks[9]. Therefore, the atomic property weighted radial distribution function (AP-RDF) is quite a good starting point, because it has a uniform length for any molecule or periodic structure. It has already been successfully applied in a number of applications with MOFs[10,11]. A similar descriptor has already been analyzed from a point of view of general ML properties, namely, the Fourier series of RDF[12]. It was shown that it has superior properties in comparison to the Coulomb matrix[12]. Namely, the Fourier series of RDF is invariant with respect to atomic indices permutations. Despite this the Coulomb matrix descriptor is more accurate by approximately 30 percent at predicting enthalpies of formation of small organic molecules[12].

To understand how the AP-RDF is related to the Coulomb matrix let us start with its definition[10]

$$RDF(R) = f \sum_{ij}^{pairs} P_i P_j e^{-B(r_{ij}-R)^2} \qquad (3)$$

Here, $f$ is the normalization constant, $B$ is the broadening parameter, $r_{ij}$ is the interatomic distance, $P_i$ is the atomic property weight, the values of $R$ are chosen based on a one dimensional grid, and the summation is running over all interatomic pairs. Since this fingerprint depends on interatomic distances and the pairwise products of atomic property weights, it is invariant with respect to translation and rotation like its Fourier series counterpart[12]. Since each grid point contains a sum of the contributions from all atomic pairs, it is also invariant with respect to atomic indices permutations. Eventually, the length of this descriptor is defined only by the number of grid points one choses to use.

A closely related descriptor, namely, an autocorrelation function[13] also belongs to the class of constant size descriptors. It was shown[13] that autocorrelation functions are more suitable for describing transition metal complexes of different sizes but with very similar spin-splitting gaps. It happens because the size truncation makes this descriptor more focused on a local chemical environment of a metal atom rather than on the overall composition of the molecule.

Further, the AP-RDF can be weighted by any physical quantity, whether it is a nuclear charge like in the Fourier series RDF[12] or, for example, hardness, electronegativity or van der Waals radii[10]. Here, we shall replace the products of atomic property weights $P_i P_j$ by atomic property matrices, $P_{ij}$. Hence, we can establish a connection to the Coulomb matrix descriptor by the following ansatz: $P_i P_j = C_{ij}$. So the Coulomb matrix elements can be uniquely transformed into positions of the AP-RDF peaks, although the width of those peaks depends on $B$. More generally, formula (3) can now be considered as a 1D representation of any two dimensional atomic centered quantities. Examples might include the Ewald sum[14] of the Coulomb matrices for periodic structures or density matrices.

One of the ways of increasing accuracy of the RDF descriptor is to augment it with angular distribution function[15–17]. However, in this work we shall combine the AP-RDF with the Bag-of-Bonds approach[3]. Note that a typical AP-RDF can contain around a hundred of points, while a diverse materials database like the Computational Ready Experimental (CoRE) database of metal-organic frameworks[18] can contain around a thousand of bond types. So even the "bags" of AP-RDF descriptor can be quite long. Fortunately, a lot of "bags" are completely empty for a diverse database with many elements from the periodic table, because the number of distinct elements in each structure is much smaller than the total number of distinct elements in the whole database.

In this situation, one of the most efficient approaches is to use kernel based methods, because the size of a kernel matrix is defined by the number of data points but not by the length of a descriptor. Indeed, the kernel is a function of two descriptor vectors $x$ and $x'$, $K(x, x')$, which outputs a scalar. Therefore, the size of the $K(x, x')$ matrix is defined by the number of $x$ and $x'$ pairs. Further, this size determines the number of model parameters that have to be optimized in such methods like support vector machine[19] and kernel ridge regression[8]. Moreover, another advantage of machine learning kernels is that they often explicitly involve a dot product of two vectors, so it is convenient to use sparse representations for the descriptors. For example, a scalar product of two vectors can be evaluated very efficiently by calculating the products of only nonzero elements[20].

The kernel approach introduces another one requirement. Namely, we cannot increase the number of data points indefinitely for kernel based methods. Therefore, it is important to have a model that does not have a tendency of overfitting data. Based on these requirements, we can now formulate the goal of this work: to design a practical descriptor that (i) would be applicable for very large and chemically diverse structures (ii) would be accurate for atomization energies (MAE ~ 1−2 kcal/mol), and (iii) would not overfit training data. The paper is organized as follows. In the next chapter we describe our formalism. In the third chapter we outline the computational details. The fourth chapter contains the results and their discussion. The fifth chapter concludes the paper with a summary.

## 2. Method

In conventional (or popular) tasks of machine learning (ML) such as image or speech recognition[9], descriptors are usually given in a raster (numeric) format, whether it is an array of pixels from an image file or time series of intensities from an audio recording. In materials informatics descriptors can easily be represented in a vector (analytic) format, because they can be mathematically derived[12] based on certain requirements. For instance, it was shown by Lilienfeld *et al.*[12] that one of the key requirements for applying descriptors in quantum mechanics is uniqueness. Based on this feature they proved a theorem which shows that a unique descriptor completely determines the corresponding quantum-mechanical observable (in the limit of infinite training data). This is similar to the connection between the external potential and the ground state density in the first Hohenberg-Kohn theorem[21], which, in other words, means that training a statistical model plays the same role as variational calculations in DFT. On the other hand, machine learning efficiently avoids variational calculations, and, as a consequence, it avoids the evaluation of the electron density.

However, the electron density is the most important concept in molecular quantum similarity[22], which provides a rigorous theoretical basis for similarity in a broader sense, and the last concept has the fundamental importance in chemistry. For example, such aspects as basicity and acidity, and even the periodic system have its roots in similarity. Therefore, in addition to the descriptor uniqueness requirement it is equally important to have a theoretical basis for molecular quantum similarity evaluation without an explicit reference to the electron density. Below we shall use machine learning kernels to make a connection between first principles kind of descriptors and quantum similarity.

Intuitively, a kernel function measures the similarity between two descriptors. So if $\langle x_i, x_j \rangle$ is a crude similarity measure between $x_i$ and $x_j$, then mapping into a new space can give a more adequate similarity function, i.e. $K(x_i, x_j) = \langle \phi(x_i), \phi(x_j) \rangle$, where $\phi(\cdot)$ is the mapping function. In practice, $\phi(\cdot)$ is not calculated explicitly. Instead, some empirical formulas satisfying Mercer's theorem[23] are employed. To demonstrate how it works, let us consider, for example, the exponentiation of the Euclidian distance between two descriptors, $d_{ij}$. This operation yields an infinite series expansion with all possible terms of the form $d_{ij}^n/n!$. So with some trial and error a nonlinearly transformed distance between descriptors can better approximate the distance between actual electron densities, i.e. $\langle \rho(x_i), \rho(x_j) \rangle$. Here, the notation $\rho(x_i)$ means that the electron density corresponds to the same molecule as the $x_i$ descriptor.

In order to get a practical outcome from such a remarkable resemblance between molecular quantum similarity and machine learning kernels let us consider several important features of the electron density. Namely, the electron density contains information about the number of electrons and the nuclear positions in the molecule which is obtained from the cusp condition[24]. Such information can also be found in molecular descriptors. Although the Coulomb matrix contains only data related to interatomic distances, combining it with other descriptors such as angular distribution functions will extend this information. Further, integrated over space, the electron density gives the number of electrons. But more important aspect is that the electron density is differentiable with respect to the number of electrons. This derivative yields the Fukui function, which is indispensable for theory of chemical reactivity[24].

It can be shown that the Coulomb matrix is also differentiable with respect to the number of electrons. Indeed, let us consider for a moment a neutral molecule. Then the number of electrons, $N_e$, equals the sum of all nuclear charges, i.e. $N_e = \sum_i Z_i$. Further, each $Z_i$ in Eqs. (1) and (2) can be expressed through the total number of electrons and the rest of nuclear charges. In the case of an ion, the total number of electrons is $N_e = \sum_i Z_i - C$, where $C$ is the ion charge. This result opens a possibility to use first principles kind of descriptors such as the Coulomb matrix, Bag-of-Bonds, radial and angular distribution functions etc. for a variety of chemical properties that are currently studied with conceptual DFT[24–26].

Thus, deriving new and more advanced first principles kind of descriptors that allow for an analytic representation can hardly be overemphasized for the sake of analysis and further improvement of machine learning kernels that can be used for molecular quantum similarity studies. In this work we consider a combination of the AP-RDF descriptor with Gaussian and Laplacian kernels.

## 2.1. Descriptor

Once we have established a connection between AP-RDF and the Coulomb matrix, let us now establish a connection with the Bag-of-Bonds, because the last descriptor performs much better than the Coulomb matrix. All interatomic distances in AP-RDF can be grouped according to a bond type $A$:

$$RDF(R) = f \sum_A \sum_{i,j \in A} P_i P_j e^{-B(r_{ij}-R)^2} \tag{4}$$

So it consists of "bags" each of which belongs to the bond type $A$ and has the properties of the total AP-RDF. However, a more important observation for our formalism is that either the total AP-RDF or each "bag" in Eq. (4) can be considered as a tensor (or strictly speaking a multidimensional array) with the first dimension spanned by the grid $R$ and the second dimension spanned by the interatomic distances $r_{ij}$, and the summation over atomic pairs corresponds to a contraction over the second dimension. These dimensions are correlated with each other through the exponential factor. Note that tensor methods are gaining a lot of attention in machine learning[27,28] and signal processing[29], because representing data in a tensor format helps to better capture such correlations between different dimensions. For example, a video is essentially a third order tensor with the third dimension being time, and highly likely that two consecutive frames in a video are correlated. We shall now use an analytic representation of AP-RDF to derive ML kernels in closed and analytic form.

## 2.2. Machine learning kernels

We shall consider two machine learning kernels that were previously employed with the Coulomb matrix based descriptors[30]: Gaussian and Laplacian. The first one is given by

$$k(x, x') = \exp\left(-\frac{1}{2\sigma^2} \|x - x'\|^2\right), \tag{5}$$

where $\|x\|$ is a Euclidean $L^2$-norm, σ is the smoothness parameter. The second one is given by

$$k(x, x') = \exp\left(-\frac{1}{2\sigma} |x - x'|\right), \tag{6}$$

where $|x|$ is a Manhattan $L^1$-norm. Note that the squared Euclidean distance can be recast as a combination of scalar products

$$\|x - x'\|^2 = xx + xx' - 2xx'. \tag{7}$$

As a result, we can introduce an overlap integral between two descriptors which will be useful for further analysis.

The starting point of our formalism is the squared Euclidean distance between two AP-RDFs which is given by

$$d_{IJ}^2 = \sum_k \left(RDF_I(R_k) - RDF_J(R_k)\right)^2 = \langle RDF_I|RDF_I\rangle + \langle RDF_J|RDF_J\rangle - 2\langle RDF_I|RDF_J\rangle, \quad (8)$$

where

$$\langle RDF_I|RDF_J\rangle = \sum_{i,j}\sum_{m,l}\int_{-\infty}^{\infty} G(R,r_{ij})G(R,r_{ml})dR, \quad (9)$$

$$G(R,r_{ij}) = fP_iP_j e^{-B(r_{ij}-R)^2}. \quad (10)$$

Here, the summations are over the indices $i, j$ of $I$ molecule and $m, l$ indices of $J$ molecule. Note that the integration in Eq. (9) is performed analytically for the exponential part of two Gaussians given by Eq. (10) according to the following formula:

$$\int_{-\infty}^{\infty} dR\, e^{-B_1(r_{ij}-R)^2} e^{-B_2(r_{ml}-R)^2} = \pi^{1/2}\gamma_{12}^{-1/2} e^{-\eta_{12}(r_{ij}-r_{ml})^2} \quad (11)$$

with $\eta_{12} = B_1 B_2/(B_1 + B_2)$ and $\gamma_{12} = B_1 + B_2$. Thus, AP-RDF does not depend on the choice of the grid spacing in this representation. Let us now analyze the product of two AP-RDFs in Eq. (9).

First of all, the product of two Gaussians can include cases when $r_{ij}$ and $r_{ml}$ belong to different bond types which leads to undesirable mixing of all molecular features. Fortunately, it is possible to reduce the number of individual Gaussian overlap integrals by calculating those only if the two interatomic distances $r_{ij}$ and $r_{ml}$ belong to the same bond type. We shall now take advantage of the AP-RDF with sorted bond types from Eq. (4) and modify the overlap integral according to the following formula:

$$\langle \widetilde{RDF}_I|\widetilde{RDF}_J\rangle = \sum_A \sum_{i,j\in A}\sum_{m,l\in A}\int_{-\infty}^{\infty} G(R,r_{ij})G(R,r_{ml})dR \quad (12)$$

Here, $A$ enumerates bond types. Thus, only the products of Gaussians with the interatomic distances of the same type contribute to the overlap integral $\langle \widetilde{RDF}_I|\widetilde{RDF}_J\rangle$. In order to formally encode this mathematical operation we introduce the complex Gaussian functions:

$$\tilde{G}(R,r_{ij},x) = e^{iZ_iZ_jx} fP_iP_j e^{-B(r_{ij}-R)^2} \quad (13)$$

So an additional integration over the auxiliary variable $x$ will incorporate the delta function:

$$\delta(Z_iZ_j - Z_mZ_l) = \frac{1}{2\pi}\int_{-\infty}^{\infty} \exp\left[ix(Z_iZ_j - Z_mZ_l)\right] dx \quad (14)$$

Due to this restriction, the Gaussian kernel with the AP-RDF descriptor will be somewhat similar to the one with Bag-of-Bonds. Let us now compare them in order to find an exact relation.

To distinguish between the Coulomb matrix elements of molecules $I$ and $J$ we shall use the following notation: $P_i P_j = C_{ij}^I$, if the pair $ij$ belongs to the molecule $I$. We shall also replace double indices by a super-index: $p \equiv ij$, $q \equiv ml$. As a result, the overlap of two bond type restricted AP-RDFs reads

$$\langle \widetilde{RDF}_I | \widetilde{RDF}_J \rangle = \sum_A \sum_{p \in A} \sum_{q \in A} f^2 C_p^I C_q^J \pi^{1/2} \gamma_{pq}^{-1/2} e^{-\eta_{pq}(r_p - r_q)^2} \quad (15)$$

On the other hand, the overlap of two Bag-of-Bonds descriptors is given by the following formula

$$\langle BoB_I | BoB_J \rangle = \sum_A \sum_{p \in A} C_p^I C_p^J \quad (16)$$

Thus, the Bag-of-Bonds overlap is constructed as a scalar product, $\vec{A} \cdot \vec{B} = \sum_i a_i b_i$, while the RDF overlap is based on the sum of all matrix elements of the tensor product, $\vec{A} \otimes \vec{B} = \sum_{i,j} a_i b_j (\vec{\alpha}_i \otimes \vec{\beta}_j)$. A direct product of two vector spaces spanned by interatomic distances allows us to use descriptors of any length without any restrictions. So we are not limited to the smallest number of interatomic distances when comparing two descriptors like in Bag-of-Bonds.

We would like to emphasize here that the main difference between our approach and other tensor kernel methods such as kernel support tensor regression[28] (STR) consists in the output format of the kernel function. Our kernel yields a scalar from two data points like in conventional methods such as support vector regression[31] (SVR), while kernel in STR yields a matrix which requires a modification of the SVR equations. At first sight it seems that we might lose some valuable information by obtaining a scalar from the summation of all elements of the tensor product. However, there is another important ingredient of our formalism. Namely, each pair of interatomic distances is multiplied in formula (15) with the corresponding exponential weight that comes from the Gaussian exponential factors when we integrate over the grid space dimension of AP-RDF in Eq. (11). That is why the correlation between two dimensions in AP-RDF mentioned above becomes important.

Note that the Bag-of-Bonds descriptor includes only a subset of all pairwise products, $C_p^I C_q^J$. It is easy to extract this subset in Eq. (15), if we apply a sorting operation for the indices $p$ and $q$ independently. This operation will not change the overlap value in Eq. (15), but it is necessary for our proof. Indeed, by comparing Eq. (15) to Eq. (16) we can now rigorously formulate the relation between two kernels. In order to obtain the Gaussian kernel with Bag-of-Bonds from the Gaussian kernel with the bond type restricted AP-RDF it is necessary to replace the tensor product in Eq. (15) by the scalar product by introducing delta function $\delta_{pq}$, so that $C_p^I C_q^J \delta_{pq} \to C_p^I C_p^J$. Then, it is necessary to eliminate the exponential factor given by Eq. (11) from each matrix element. As such the Bag-of-Bonds descriptor is a subset of the bond type restricted AP-RDF.

As for the application of the Laplacian kernel given by Eq. (6) to AP-RDF, we cannot use Eq. (11) for the analytical evaluation. Therefore, to introduce the bond type restriction operation we have modified Manhattan distance according to the following formula:

$$d_{IJ} = \sum_k \sqrt{RDF_I(R_k)RDF_I(R_k) + RDF_J(R_k)RDF_J(R_k) - 2RDF_I(R_k)RDF_J(R_k)} \quad (17)$$

So the bond type restriction is applied at every grid point $R_k$ numerically

$$RDF_I(R_k)RDF_J(R_k) = \sum_A \sum_{i,j \in A} \sum_{m,l \in A} G(R_k, r_{ij}) G(R_k, r_{ml}) \quad (18)$$

We shall now describe computational details of our method.

## 3. Computational Details

We have implemented all descriptors and kernels using the Python programming language with the *NumPy* package[32] to utilize its C backend for efficiency. There is only one exception: the overlap integral between Gaussian functions given by Eq. (11) has been implemented directly in C language. We have obtained the regression coefficients from the Kernel Ridge Regression[8] (KRR) method. We have used QM7 benchmark database from Reference 1. This database consists of geometries and atomization energies calculated with DFT of 7165 molecules. Each molecule contains up to 23 atoms such as C, N, O, S, and H. This database also contains five subsets of indices of all 7165 molecules for the five-fold cross-validation, which we have also employed in our statistical modelling. We have used random search approach[33] combined with genetic algorithm[34] for finding best values of hyper-parameters of the model.

## 4. Results and Discussion

To eliminate possible discrepancies due to the use of different training sets, validation techniques and other issues, we have trained our new models as well as those that were studied earlier by making use of exactly the same validation scheme and the same benchmark set. Namely, we have used atomization energies from the QM7 molecular benchmark set and KRR method, because this combination has been used earlier. The optimal values of hyper parameters such as σ (kernel parameter), λ (regularization parameter) and *B* (broadening of RDFs) have been obtained from the five-fold cross-validation procedure according to Reference 1. The results in the form of mean absolute errors (MAE) are given in Table 1. First we shall compare results for the Gaussian kernel.

**Table 1**. Five-fold cross validation MAE (kcal/mol) for different models obtained with KRR; QM7 set

| Descriptor | Gaussian | Laplacian |
|---|---|---|
| **Bond type restricted RDF** | 1.7 | 3.6 |
| **Bag-of-Bonds** | 5.1 | 2.4 |
| **Sorted Coulomb matrix** | 8.6 | 4.3 |
| **RDF** | 9.4 | 11.2 |
| **Eigenvalues of Coulomb matrix** | 10 | 9.8 |

When we compare the bond type restricted AP-RDF against the Bag-of-Bonds result, we find that our descriptor is three times more accurate (1.7 against 5.1 kcal/mol). We shall investigate numerically the origin of such an improvement later. Further, the effect of the bond type restriction yields more than five-fold improvement when we compare the bond type restricted AP-RDF to the original AP-RDF (1.7 against 9.4 kcal/mol). It is interesting to note that this effect is not so pronounced when we compare Bag-of-Bonds to the sorted Coulomb matrix (5.1 against 8.6 kcal/mol). On the other hand, the original AP-RDF performs slightly worse than sorted Coulomb matrix, and this is in agreement with previous studies on the Fourier series RDF[12].

As for the differences between the results from Gaussian and Laplacian kernels, neither of two kernels performs better for all descriptors (see Table 1). It was already reported in the literature that for the eigenvalues of the Coulomb matrix the performance is very similar. The Fourier series RDF[12] yields smaller MAE with the Gaussian kernel, which is in agreement with our results for the original AP-RDF. As for the sorted Coulomb matrix and Bag-of-Bonds, both of them yield approximately a two-fold improvement when the Gaussian kernel is replaced by the Laplacian kernel. Therefore, it is quite surprizing that the bond type restricted AP-RDF yields twice less accurate results with the Laplacian kernel than with the Gaussian one (3.6 against 1.7 kcal/mol). We shall now analyze the results for the bond type restricted AP-RDF in more details.

First, we shall compare the bond type restricted AP-RDF and Bag-of-Bonds within the Gaussian kernel to explain the source of the three-fold improvement in MAE. As we have already mentioned in the Method section, the Gaussian kernel contains the squared Euclidian distance between two vectors, and the distance is expressed in terms of the overlaps. So we shall analyze numerically the overlap formulas obtained in analytic form, i.e. Eqs. (15) and (16). To be concrete let us now consider as an example the overlap matrix between methane and ethanol based on Eqs. (15) and (16). Table 2 demonstrates that a general structure of the overlap either for two Bag-of-Bonds or two bond type restricted AP-RDFs is the same, because both methods employ the bond type restriction. Therefore, the total overlap matrix contains different blocks. This is highlighted in Table 2 by dashes when there is no overlap and by **X** when there is a nonzero overlap. By the way, Table 2 also highlights that in the kernel based approach there is no need to store a lot of null matrix elements for both descriptors in computer memory, for instance, OC and OH bags of methane are never used, which can be critical for very large systems.

**Table 2**. The structure of the product of two descriptors, i.e. two Bag-of-Bonds or two bond type restricted RDFs. The rows correspond to the bond types in methane and columns to ethanol

|    | OC | OH | CC | CH | HH |
|----|----|----|----|----|----|
| CC | –  | –  | $\mathbf{X}_{CC}$ | – | – |
| CH | –  | –  | –  | $\mathbf{X}_{CH}$ | – |
| HH | –  | –  | –  | –  | $\mathbf{X}_{HH}$ |

We shall now consider a nonzero overlap sub-matrix $\mathbf{X}_{CH}$ between two CH "bags" for both descriptors. There are four CH bonds in methane, and all of them are equal, so there is only one subgroup of the Coulomb matrix elements. In ethanol there are twelve CH interatomic distances which can be placed into four subgroups according to their values. The first subgroup contains five covalent CH bonds. The next subgroup also contains five components formed by the interatomic distances between the hydrogens of CH bonds and non-neighbour carbon atoms. Eventually, the last two subgroups contain one component each formed by the interatomic distances between the hydrogen of the OH bond and two carbon atoms. We emphasize here that in the Bag-of-Bonds overlap sub-matrix $\mathbf{X}_{CH}$ each Coulomb matrix element of methane is multiplied by the largest value from the ethanol "bag". These largest matrix elements correspond to the smallest values of interatomic distances in the CH bonds of ethanol. As a result, there are four equal matrix elements in the CH overlap sub-matrix for the Bag-of-Bonds, as we can see in the top image of Figure 1. On the other hand, there are four distinct groups of matrix elements for the CH overlap sub-matrix for the bond restricted RDF (see the bottom image of Figure 1).

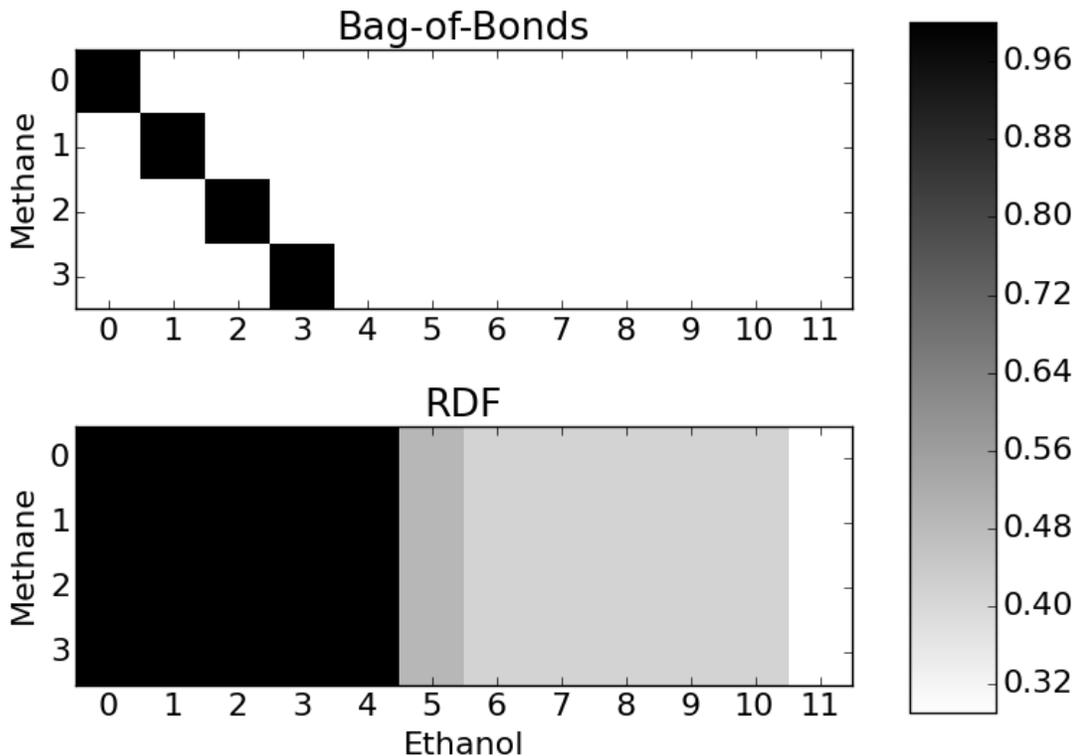

**Figure 1.** Shown is the overlap matrix for a single CH bag in Bag-of-Bonds and the bond type restricted AP-RDF. The matrix elements are normalized to be less than one. The heat map is based on the numerical values of the matrix elements. The axis labels enumerate the bond distances.

Note that although we have used quite a small value of the broadening parameter ($B = 0.1$), Figure 1 correctly describes the dissimilarity between RDF and Bag-of-Bonds overlaps, albeit qualitatively. However, there is an additional nuance that should be taken into account. The matrix elements of the RDF overlap matrix include not only the product of two Coulomb matrix elements, $X_{ij,ml} = C_{ij}C_{ml}$, but they also include an extra factor given by Eq. (11) which inversely depends on the difference between interatomic distances, i.e. $\tilde{X}_{ij,ml} = X_{ij,ml}\pi^{1/2}\gamma_{pq}^{-1/2}e^{-\eta_{pq}(r_{ij}-r_{ml})^2}$. So when the interatomic distances differ very much, then the corresponding matrix element is substantially damped. To analyze this effect we have shown in Figure 2 a distribution of the overlap values between ethanol and the largest molecules from the QM7 database that contain more than 20 atoms. In this case we have used a realistic value of the broadening parameter ($B = 19$, see Table 3). We remind the reader here that the Bag-of-Bonds overlap is the sum of only diagonal matrix elements, while the RDF overlap is the sum of all matrix elements. As we can see in Figure 2, the majority of the values of the Bag-of-Bonds overlap are between 71.5 and 72 in the units of the Coulomb matrices product. On the other hand, there is a differentiation in the values of the bond restricted AP-RDF overlap that ranges from 160 to 280 in the same units.

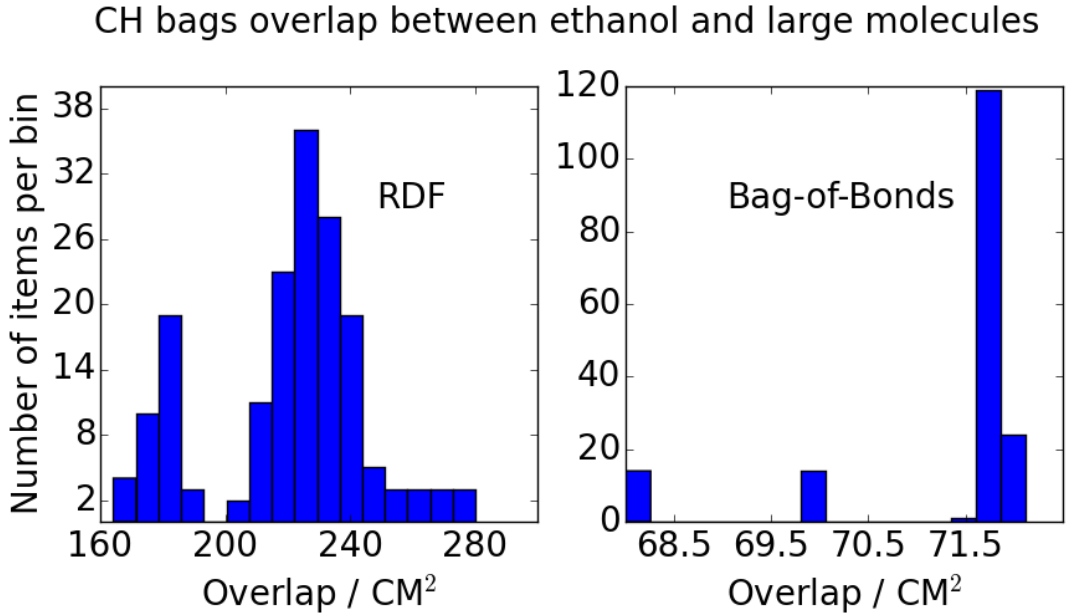

**Figure 2.** Shown is the distribution of the overlap values for CH bags between ethanol and molecules with more than 20 atoms from QM7 database.

**Table 3.** The results for models based on bond type restricted RDF and KRR; QM7 set

| | Bond type restricted RDF | | | |
|---|---|---|---|---|
| **Kernel** | λ | σ | *B* | MAE (kcal/mol) |
| **Gaussian** | $10^{-6}$ | 167 | 19 | 1.8[a] |
| **Laplacian** | $10^{-6}$ | 4418 | 2.8 | 3.6 |

[a] A slightly more accurate number (1.7 kcal/mol) was obtained with the optimized normalization factor *f* in Eq. (4) and including the diagonal Coulomb matrix elements into AP-RDF.

This feature of the RDF overlap allows us to compare molecules with different sizes more efficiently than with Bag-of-Bonds. To visualise this difference even better we have shown in Figure 3 Euclidean distances for both descriptors calculated between 98 smallest and 172 largest molecules from the QM7 database. The smallest molecules contain less than 10 atoms, while the largest ones contain more than 20. The values of the matrix elements are scaled to be less than one for comparison purposes. As we can see from Figure 3, there is more variation in the values of the matrix elements for the bond restricted RDF than for Bag-of-Bonds. In other words, the RDF distances are very specific for each molecular pair, while Bag-of-Bonds yields distances either between 0.1 and 0.2 or between 0.9 and 1.0. Thus, our statistical model can potentially be more transferable for databases that include large variation in sizes of molecules.

All this numerical analysis corroborates our findings from the Method section about the advantages of using the tensor product instead of the scalar product in the kernel. However, that analysis is relevant for the Gaussian kernel. We shall now investigate why the Laplacian kernel improves MAE for the Bag-of-Bonds, but it gives rise to an opposite effect for the bond type restricted AP-RDF.

First of all, the results in Table 1 for the AP-RDF descriptors with the Gaussian kernel have been obtained by making use of analytical integral given by Eq. (15), while the results for the Laplacian kernel have been obtained by making use of numerical integration with Eq. (18). To eliminate possible discrepancies due to analytical and numerical schemes we have estimated numerical grid values, $\Delta R$, for accurate numerical representation of the AP-RDF. The details are given in the Appendix. Since we have used a fine grid for obtaining the results in Table 1, the results for the Laplacian kernel are accurate.

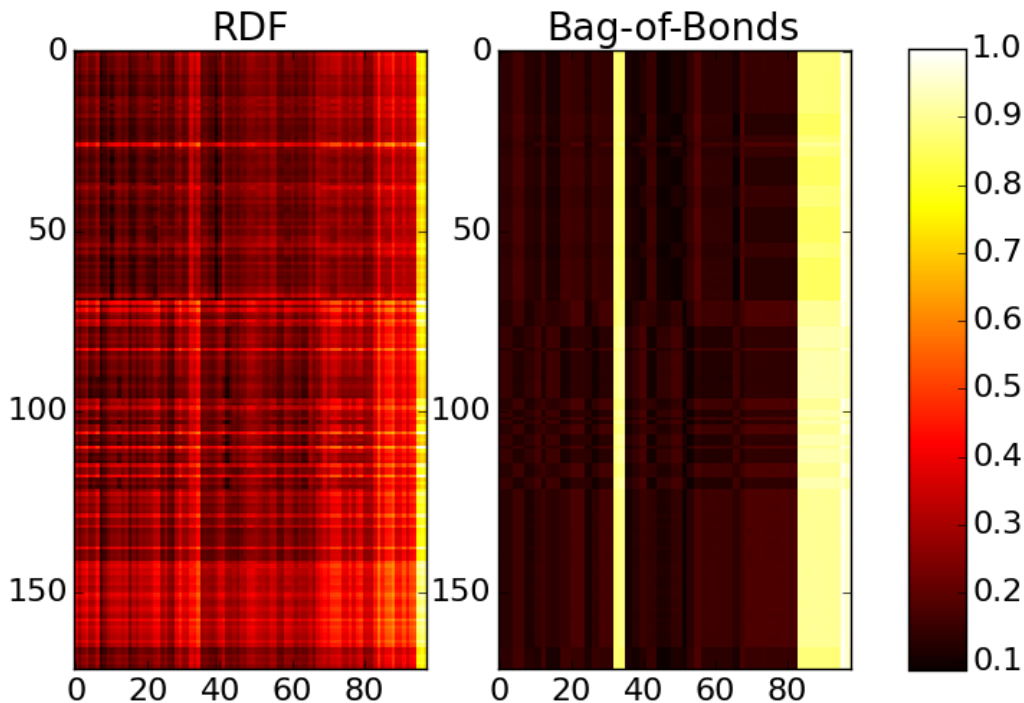

**Figure 3.** Euclidean distances between 98 smallest (less than 10 atoms) and 172 largest (more than 20 atoms) molecules; the value of parameter $B = 19$ for bond restricted RDF. The overlap values are scaled to one for convenience.

We think that an increase of the validation errors for AP-RDFs when using the Laplacian kernel can be attributed to the overfitting of the training data, because the bond type restriction and Laplacian kernel both increase "flexibility" of the statistical model. By "flexibility" we mean here the ability to reproduce training values. See, for example, a comparison of two kernels in Reference [30]. To test this hypothesis we have compared not only the cross-validation errors for different models and kernels but the corresponding training errors as well, i.e. the error of the model on the training set only. The results of our calculations are summarized in Table 4. As we can see from the table, the level of overfitting by the bond type restricted AP-RDF using the Laplacian kernel is very high, such that the training error is approximately one thousand times smaller than the cross-validation error. Note that the overfitting is less severe for the Bag-of-Bonds descriptor with the Laplacian kernel. Nevertheless, the training error is smaller than the cross-validation error by two orders of magnitude. Also note that the Gaussian kernel does not suffer from the overfitting problem for any descriptor. Therefore, it is more suitable for our initial goals.

**Table 4**. Training and cross-validation MAE for different models obtained with KRR; QM7 set

| Kernel | Bond type restricted RDF | | Bag-of-Bonds | |
|---|---|---|---|---|
| | Training error (kcal/mol) | Cross-validation error (kcal/mol) | Training error (kcal/mol) | Cross-validation error (kcal/mol) |
| **Gaussian** | 0.2 | 1.7 | 0.2 | 5.1 |
| **Laplacian** | $10^{-3}$ | 3.6 | $10^{-2}$ | 2.4 |

Here, we should mention the discrepancies between the Bag-of-Bonds results obtained with our code which are shown in Table 1 and the Bag-of-Bonds results reported by Hansen *et al.* in Reference 3. More specifically, using the QM7 database Hansen *et al.* report a MAE of 1.5 kcal/mol for the Laplacian kernel, while we report a MAE of 2.4 kcal/mol. Similarly for the Gaussian kernel Hansen *et al.* report a MAE of 4.5 kcal/mol, while we report a value of 5.1 kcal/mol. These discrepancies arise from two sources. First, we have used the five-fold cross-validation, while the results from Reference 3 were obtained with a simple out-of-sample validation scheme. Second, we have used the original QM7 benchmark database from Reference 1, while the results from Reference 3 were obtained with a smaller QM7 database from Reference [35] with 7101 molecules instead of 7165. Although we managed to reproduce the results of Reference 3 with the modified database from Reference [35], we think that more reliable and revealing results are obtained with the five-fold cross-validation, i.e. 2.4 instead of 1.5 kcal/mol for the Laplacian kernel and 5.1 instead of 4.5 kcal/mol for the Gaussian kernel.

We shall now discuss the length of descriptors issue mentioned in the Introduction. By "length" we mean the number of features or columns of the descriptor, for example, the number of eigenvalues or the number of interatomic distances. Numerical AP-RDF has a uniform length defined by the limits of the numerical grid. The same is true for each "bag" of the bond type restricted AP-RDF, while the Bag-of-Bonds size varies quadratically with the number of atoms $n$ in each "bag" as $n(n-1)/2$. Note that the size of analytic AP-RDF also grows quadratically, because we have to store parameters of all atomic pairs. In order to make the Bag-of-Bonds descriptor more manageable in size, one could not include pairs that are greater than a given cut-off distance, $R_C$. Using such a truncation scheme, let us consider the Bag-of-Bonds descriptor for a hypothetical linear chain of atoms where the bond distances are all 1 Å in length. Figure 4 shows the size of the Bag-of-Bonds descriptor for this hypothetical system as a function of the cut-off distance, $R_C$. This shows that the Bag-of-Bonds descriptor will continue to grow even with a cut-off. On the other hand, numeric AP-RDF is a constant size descriptor. So numeric bond type restricted AP-RDF would be the best choice for very large systems both in terms of size and accuracy.

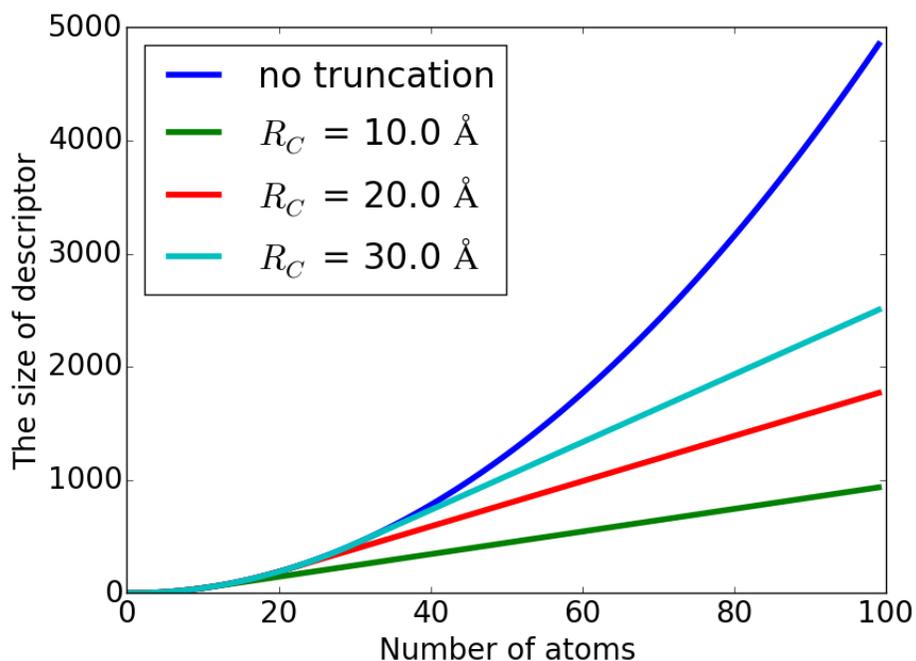

**Figure 4.** The number of interatomic distances in the Bag-of-Bonds (size) as a function of the number of atoms in a linear chain.

## 5. Conclusion

We have implemented a new machine learning formalism which efficiently employs sparse representation for a descriptor and a tensor product inside of machine learning kernels. The descriptor is based on the atomic property weighted RDF which uses properties derived from the Coulomb matrix. However, this approach avoids the problem of other Coulomb matrix based descriptors, namely, the calculation of very long descriptor vectors for very large molecular systems. Moreover, our approach yields MAE in the range 1−2 kcal/mol for atomization energies of small organic molecules, which is lower than MAE of all other descriptors of this kind. Such high level of accuracy is achieved by the bond type restriction in the AP-RDF. Due to this restriction our model is five times more accurate than the one based on a simple AP-RDF without such a restriction.

The idea of bond type restriction was first implemented in Bag-of-Bonds[3], but our formalism is more general and more accurate. The generalization is achieved due to a tensor product, which allows us to compare descriptors of different lengths. As for efficiency, our model allows for both an analytic or numerical representation. While the analytic form may be more elegant, the numerical form is much more computationally efficient. The latter consists of the calculation of as many RDFs as there are bond types in a given structure. We have shown that it can also be as accurate as the analytic form with a careful choice of numerical grid parameters. We have estimated the values of such parameters from analytical integration.

It should be mentioned here that there are more accurate descriptors than the Coulomb matrix based ones[4], but our model also allows for an extension. For example, it is possible to include additional information like angles between chemical bonds in the form of angular distribution function[15–17] or dihedrals[5]. However, we have explored in this work advantages and limitations of a simpler descriptor like AP-RDF to utilize its full potential. And although we have tested it only together with KRR, but it can be employed with other kernel based methods such as support vector regression[19]. In this regard, it was discussed how the mapping function in machine learning kernels is related to electron density when first principles kind of descriptors are used. However, our formalism is not restricted to the kernel based methods, and it can also be extended to radial basis function networks[36]. As for other extensions, this is the first application of our formalism to molecular systems, and we shall apply it later to metal-organic frameworks in subsequent publications.

## Appendix

We have evaluated the accuracy of numerical integration in the original AP-RDFs, i.e. AP-RDFs without bond type restriction given by Eq. (3). For this purpose we have optimized the hyper parameters of both analytic and numeric AP-RDFs on the same molecular training/validation set as before. The values of hyper parameters together with MAE are given in Table 5. As we can see from the table, both approaches yield the same predictions, and the optimal value of the grid spacing (0.077 Bohr or 0.04 Å) is quite small. Note also that the value of the broadening parameter $B$ is quite high. Therefore, the individual Gaussian functions (10) are very sharp, so it is not surprizing that the grid spacing has to be quite small in this situation.

**Table 5.** The results for models based on AP-RDFs with Gaussian kernel and KRR; QM7

| RDF type | λ | σ | $B$ | $\Delta R$ (Bohr) | MAE (kcal/mol) |
|---|---|---|---|---|---|
| analytic | $10^{-4}$ | 33 | 88 | – | 9.4 |
| numeric | $10^{-4}$ | 35 | 79 | 0.077 | 9.4 |